\documentclass[twocolumn]{aastex7}
\usepackage{xspace,amsmath,amssymb,multirow}

\newcommand{\per}{\ensuremath{^{-1}}\xspace}

\newcommand{\Lya}{Ly\ensuremath{\alpha}\xspace}

\shorttitle{}
\shortauthors{Taylor et al.}

\begin{document}

\title{Spectroscopic Ultraluminous Ly$\mathbf{\alpha}$ Luminosity Functions at $\mathbf{z=5.7}$ and $\mathbf{z=6.6}$ from HEROES: Evidence for Ionized Bubbles}

\correspondingauthor{Anthony~J.~Taylor}
\email{anthony.taylor@austin.utexas.edu}

\author[0000-0003-1282-7454]{A.~J.~Taylor}
\affiliation{Department of Astronomy, The University of Texas at Austin, Austin, TX, USA}
\affiliation{Department of Astronomy, University of Wisconsin-Madison,
475 N. Charter Street, Madison, WI 53706, USA}
\email{anthony.taylor@austin.utexas.edu}

\author[0000-0002-3306-1606]{A.~J.~Barger}
\affiliation{Department of Astronomy, University of Wisconsin-Madison,
475 N. Charter Street, Madison, WI 53706, USA}
\affiliation{Institute for Astronomy, University of Hawaii, 2680 Woodlawn Drive,
Honolulu, HI 96822, USA}
\affiliation{Department of Physics and Astronomy, University of Hawaii,
2505 Correa Road, Honolulu, HI 96822, USA}
\email{barger@astro.wisc.edu}

\author[0000-0002-6319-1575]{L.~L.~Cowie}
\affiliation{Institute for Astronomy, University of Hawaii,
2680 Woodlawn Drive, Honolulu, HI 96822, USA}
\email{cowie@ifa.hawaii.edu}

\author[0009-0008-7427-4617]{E.~M.~Hu}
\affiliation{Institute for Astronomy, University of Hawaii,
2680 Woodlawn Drive, Honolulu, HI 96822, USA}
\email{hu@ifa.hawaii.edu}

\author[0000-0001-9028-6978]{A.~Songaila}
\affiliation{Institute for Astronomy, University of Hawaii,
2680 Woodlawn Drive, Honolulu, HI 96822, USA}
\email{acowie@ifa.hawaii.edu}

\begin{abstract}
We present spectroscopic \Lya luminosity functions (LFs) at $z=5.7$ and $z=6.6$ based on a large 209 source sample of \Lya emitter (LAE) candidates identified in Subaru/Hyper Suprime-Cam narrowband imaging and confirmed with Keck~II/DEIMOS spectroscopy over a multi-year observing campaign. After applying photometric and spectroscopic cuts to produce homogeneous samples, we use the resulting samples of 49 $z=5.7$ and 56 $z=6.6$ LAEs to compute spectroscopic LAE LFs at each redshift. We correct our LFs for incompleteness using a source-injection simulation. We find excellent agreement with current spectroscopic and photometric LAE LFs from the literature. We look for evolution over the redshift range $z=5.7$--6.6. We find a strong convergence of the LFs at $L_{\textrm{\Lya}}\gtrsim 10^{43.4}~\textrm{erg~s}^{-1}$. This convergence (noted in previous literature) provides strong evidence that the most luminous LAEs at $z=6.6$ form ionized bubbles around themselves, allowing for greater \Lya transmission through the neutral intergalactic medium, which is measured as an increase in the bright end of the $z=6.6$ LF over the expected evolution in the LF based on the faint end. We infer that ultraluminous LAEs may play a significant role in reionization. 
\end{abstract}

\section{Introduction} \label{sec:intro}

Determining the redshift and timeline of reionization, as well as the sources of the photons that reionized the intergalactic medium (IGM), are key goals in observational cosmology. At high redshifts ($z>5$), rest-frame optical emission lines, such as H$\alpha$, H$\beta$, [O\,\textsc{iii}]$\lambda$5007, and [N\,\textsc{ii}], are redshifted into near-infrared (NIR) wavelengths inaccessible from the ground, while the \Lya emission line and rest-frame UV continuum are redshifted into the optical/NIR. Early, powerfully star-forming Ly$\alpha$ emitters (LAEs) in this epoch can therefore be studied from the ground. 

The \Lya emission line is an ideal probe of galaxies and their environments, as the strength and shape of the line is modified by the radiative damping wings of neutral hydrogen in the IGM \citep[e.g.][]{gronke20}. 
As the neutral hydrogen fraction in the IGM increases with redshift, \Lya lines become narrower and less luminous, and only the red wings of the lines remain visible \citep[e.g.,][]{songaila22,songaila24}. At the highest redshifts, we may no longer see \Lya lines at all. 

However, ultraluminous LAEs (ULLAEs) may remain visible during the epoch of reionization by generating giant ionized bubbles around themselves, possibly in association with neighboring galaxies \citep{hu16,santos16,matthee18,tilvi20,meyer21,songaila22,songaila24}. In this case, the neutral hydrogen scattering by the IGM does not take place, and the intrinsic profiles of the \Lya lines---which in some cases can have both blue and red components---are visible. Then, through theoretical modeling, quantities like the escape fraction of ionizing photons from the parent galaxy, which are critical to understanding the ionization process, can be inferred \citep[see, e.g.,][]{gronke20}.

Searches for intrinsically rare ULLAEs require large-area surveys. In recent years, instruments such as the Hyper Suprime-Cam (HSC) imager on the Subaru 8.2m telescope have enabled such surveys, due to its large aperture coupled with its wide 1.5~deg diameter field-of-view. In particular, narrowband surveys have the potential to identify efficiently $\sim$hundreds of $z>5$ LAEs across $\sim$tens of square degrees. The NB921 and NB816 narrowband filters (centered at 9210~\AA{} and 8160~\AA{}, respectively) are particularly useful for these searches, as they are sensitive to \Lya emission at $z=6.6$ and $z=5.7$, respectively, conveniently probing both during and immediately after the epoch of reionization. Indeed, the SILVERRUSH program \citep{konno18,umeda25} has used these filters with the HSC Strategic Subaru Program \citep[HSC-SSP;][]{aihara18,aihara19,aihara22} to select $\sim6600$ photometric LAE candidates at $z>5$ over $24.4$~deg$^2$ of imaging. Similarly, the LAGER program has used the DECam instrument on the 4m Blanco Telescope to produce samples of $z=6.93$ LAE candidates using a NB964 filter \citep{wold22}.

However, photometric selection of high-redshift LAEs can be affected by artifacts in the data and by low-redshift contaminants, such as [O\textsc{iii}]$\lambda$5007 emitters at $z\sim0.6-0.9$ \citep{konno18,taylor20,taylor21}. Thus, spectroscopic confirmation of the \Lya line, especially in the intrinsically rare ultraluminous regime, is crucial for ensuring pure LAE samples. 

Large spectroscopic follow-up programs using multi-object spectrographs are needed to constrain the ultraluminous end of the LAE Luminosity Function (LF). For example, \citet{ning22} used the M2FS fiber spectrograph on the Magellan 6.5m telescope to follow up HSC-selected $z=5.7$ and $z=6.6$ LAE photometric candidates to produce spectroscopic LAE LFs. Similarly, we used the DEIMOS spectrograph on the Keck~II 10m telescope to follow up LAE candidates selected from the $\sim$45~deg$^2$ HEROES \citep[Hawaii EROsita Ecliptic pole Survey;][]{taylor23}. By adopting a ``brightest-first'' observational strategy, we constructed ultraluminous LAE LFs at $z=6.6$ and $z=5.7$ \citep{taylor20,taylor21}. We now update these LFs with our latest spectroscopic LAE samples (as first introduced in \citealt{songaila24}), and we analyze the largest spectroscopic sample of $z=5.7-6.6$ ULLAEs to date. 

This article is organized as follows. In Section~\ref{sec:observations}, we describe our photometric and spectroscopic observations and our resulting LAE samples. In Section~\ref{sec:lum}, we calculate the \Lya line fluxes and luminosities, and we present our LAE catalogs. In Section~\ref{sec:completeness}, we detail our incompleteness corrections and spectroscopic completeness in preparation for our construction and analysis of the $z=5.7$ and $z=6.6$ LAE LFs in Section~\ref{sec:LFs}. Finally, in Section~\ref{sec:summary}, we summarize our results and describe our future plans.

We assume $\Omega_M$=$0.3$, $\Omega_\Lambda$=$0.7$, and $H_0$=$70$~km~s$^{-1}$~Mpc$^{-1}$ throughout. All magnitudes are given in the AB magnitude system, where an AB magnitude is defined by $m_{AB}$=$-2.5\log f_\nu - 48.60$. Here $f_\nu$ is the flux of the source in units of ergs~cm$^{-2}$~s$^{-1}$~Hz$^{-1}$. We use `$\log$' to refer to the base-10 logarithm throughout.

\begin{deluxetable*}{cccccccc}
\tablewidth{0pt}
\tablecaption{Summary of Observed Fields}
\label{tab:fields}
\tablehead{Field & Area & Comoving Volume & Comoving Volume & LAEs &  LAEs & NB816 & NB921 \cr
& deg$^2$ & $\times10^6$~Mpc$^3$ (z=5.7) & $\times10^6$~Mpc$^3$ (z=6.6) & (z=5.7) & (z=6.6) & (5$\sigma$ mag) & (5$\sigma$ mag)}
\startdata
NEP (excluding TDF) & 41.01 & 31.07 & 36.60 & 29 & 31 & 24.66 & 24.63\\
XMM-LSS & 8.29 & 6.28 & 7.40 & 7 & 3 & 25.57 & 25.53\\
SSA22 & 5.74 & 4.35 & 5.12 & 11 & 2 & 25.36 & 24.88\\
TDF & 2.71 & 2.05 & 2.42 & 2 & 17 & 24.98 & 25.29\\
COSMOS & 1.76 & 1.33 & 1.57 & 0 & 3 & 26.37 & 26.24\\ \hline
Total & 67.8 & 45.08 & 53.11 & 49 & 56 & \nodata & \nodata
\enddata
\end{deluxetable*}

\section{Observations}\label{sec:observations}

We adopt the uniformly observed samples of 136 LAEs at $z\sim5.7$ and 84 LAEs at $z\sim6.6$ from \citet{songaila24}, where full details of the observations are given. Briefly, we selected LAEs from large Subaru/HSC and Subaru/SC imaging surveys in the North Ecliptic Pole (NEP), SSA22, COSMOS, and XMM-LSS fields. The NEP was our primary field of interest as the target of our HEROES program. HEROES is a 45~deg$^{2}$ Subaru/HSC broadband and narrowband survey (using the HSC-G, HSC-R2, HSC-I2, HSC-Z, HSC-Y, NB0816, and NB0921 filters, referred to hereafter as the $g$, $r$, $i$, $z$, $y$, NB816, and NB921 filters) that was principally designed to enable LAE selections (see \citealt{taylor23} for details). In a subset of the NEP field, \cite{taylor23} obtained deeper HSC imaging in NB921, $z$, and $r$ to produce higher quality optical imaging over the JWST Time Domain Field \citep[TDF,][]{windhorst17,windhorst23}. Due to the differences in imaging depth, we treat the TDF and the broader NEP separately throughout this study. We also observed the SSA22 field using the same strategy, and we used the publicly available Deep/UltraDeep HSC imaging in the XMM-LSS and COSMOS fields from the HSC-SSP \citep{aihara18,aihara19,aihara22}.

In addition, we observed LAEs in the Hubble Deep Field North (HDF-N), A370, GOODS-N, SSA13, SSA17, and Lockman Hole fields, but due to the small number of observed sources, we exclude these fields from our analysis to greatly simplify our sample completeness estimates (see \S\ref{sec:completeness}). The samples we use are the summation of an $\sim$15 year observing effort. Our source selection criteria have changed over time, depending on the length of observing runs, available imaging filters, ever increasing imaging depths, and evolving project goals. 

\begin{figure*}[htp]
\centering
\includegraphics[width=\textwidth]{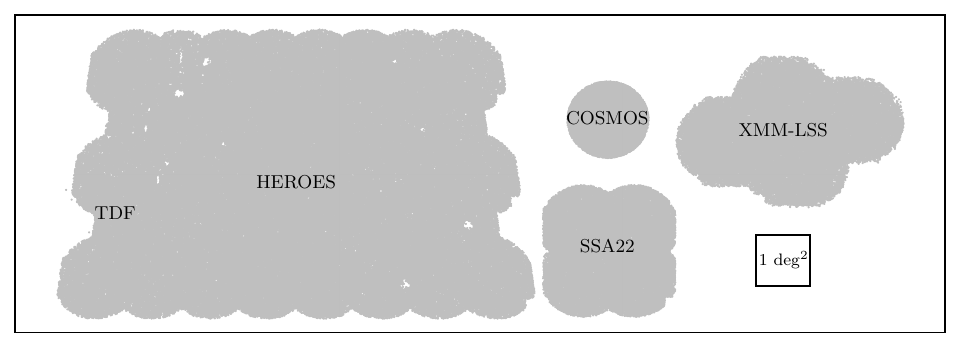}
\includegraphics[width=0.49\textwidth]{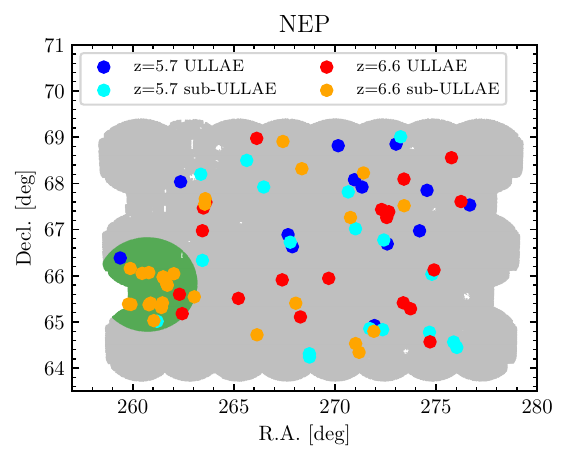}
\includegraphics[width=0.49\textwidth]{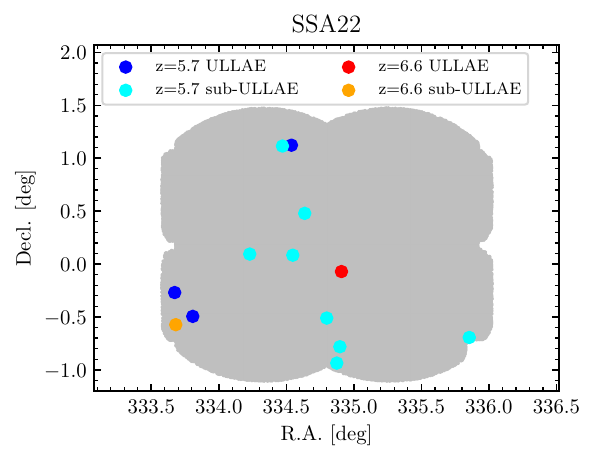}
\includegraphics[width=0.49\textwidth]{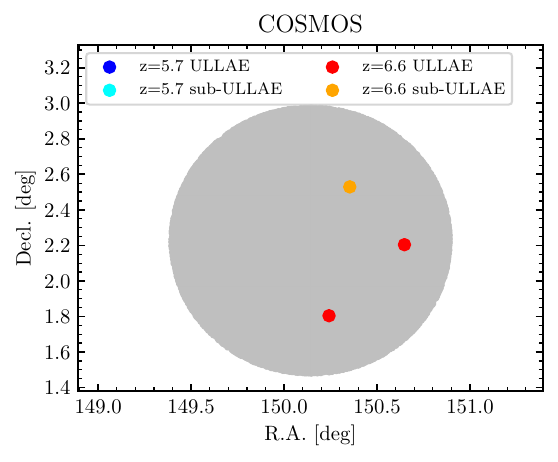}
\includegraphics[width=0.49\textwidth]{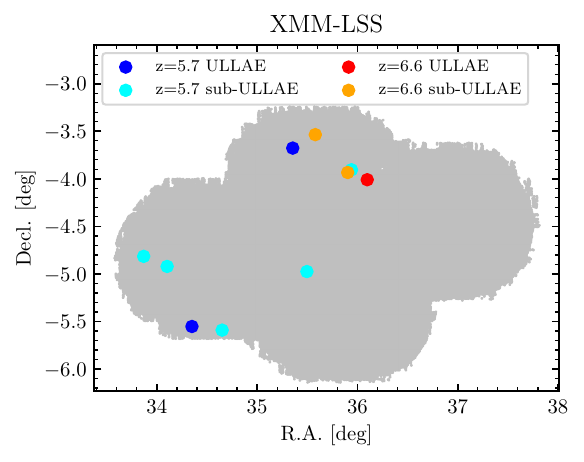}
\caption{Top: The four fields used for the LF LAE samples reprojected to a common equal-area scale to demonstrate the relative areas of the fields. The black square shows a 1~degree $\times$ 1~degree region for scale. Bottom panels: The LAE samples in each field. The field areas are shaded gray, the ULLAEs (with $L_{\textrm{\Lya}}>10^{43.5}$~erg~s$^{-1}$) at $z=5.7$/$z=6.6$ are shown as blue/red circles, and sub-ULLAEs at $z=5.7$/$z=6.6$ (with $L_{\textrm{\Lya}}<10^{43.5}$~erg~s$^{-1}$) are shown as cyan/orange circles. The green shading indicates the area of increased imaging depth centered on the JWST-TDF in HEROES.}
\label{fig:radec}
\end{figure*}

However, to understand the completeness of our samples to construct \Lya LFs, we require homogeneously selected samples. To best capture the majority of our observed objects while adopting a common set of photometric selection criteria, in this work, we use the following: For the $z=5.7$ LAEs, we require signal-to-noise (S/N)$<5$ non-detections in $g$ and $r$, magnitudes $20<\textrm{NB816}<24.5$, $i-\textrm{NB816}>0.7$, and $z-\textrm{NB816}>0.7$. For the $z=6.6$ LAEs, we require S/N$<5$ non-detections in $g$, $r$, and $i$, magnitudes $20<\textrm{NB921}<24.5$, and $z-\textrm{NB921}>0.7$.  These cuts generally enforce a strong \Lya break, a significant narrowband excess over neighboring broadbands, and narrowband brightness cuts so as to select viable candidates for spectroscopic detection. These narrowband brightness cuts are also brighter than the 5$\sigma$ narrowband depths (measured in 2'' diameter apertures) in all of our selected fields (see Table~\ref{tab:fields}). We apply these cuts uniformly to all of the fields, except for the TDF, in which we relax the NB921 brightness cut to $20<\textrm{NB921}<25.0$ to best include data from observing runs that specifically targeted fainter LAEs in the TDF. Finally, we require that the measured spectroscopic redshifts place the \Lya line where the narrowband filter transmission is at $>$50\% of the peak transmission (8118--8220~\AA{}, $\boldsymbol{[5.678<z<5.762]}$ for NB816, and 9137--9270~\AA{}, $\boldsymbol{[6.516<z<6.625]}$ for NB921). This cut is important for reasonable flux calibration, as we discuss in Section~\ref{sec:lum}. We retroactively apply these criteria to our observed samples to produce final homogeneously selected samples consisting of 49 $z\sim5.7$ LAEs and 56 $z\sim6.6$ LAEs. In Table~\ref{tab:fields}, we list the fields, surveyed areas, probed comoving volumes (calculated in Section~\ref{sec:LFs}), and LAE counts. In Figure~\ref{fig:radec}, we show the surveyed areas and the locations of the LAEs within these areas.

At lower redshifts ($z<5$), quasars significantly contaminate the bright end of the \Lya LF \citep[e.g.,][]{umeda25}. However, our LAE samples are free of AGN contaminants due to selection cuts applied in \cite{songaila24}. In brief, \cite{songaila24} tested for the presence of AGN by (1) examining their sample of LAE spectra for N\,\textsc{v}~$\lambda\lambda$1239,1243 emission, and (2) searching for significantly broad \Lya lines (FWHM$\gtrsim$1000~km~s$^{-1}$). \cite{songaila24} removed three AGN from their sample using these tests. Thus, as we adopt the final samples from \cite{songaila24}, our sample is free from broad-line and N\,\textsc{v}~$\lambda\lambda$1239,1243 emitting AGN contaminants. 

\section{Line Fluxes and Luminosities}
\label{sec:lum}
We next calculate \Lya line fluxes and luminosities for these final LAE samples. Following our previous methodology in \cite{taylor20,taylor21}, in lieu of performing a challenging flux calibration from the spectroscopic data subject to potential slit-losses and other spectrograph instrumental effects, we instead use the high S/N measurements of the LAEs from the narrowband imaging. Here, we assume that the \Lya line is the dominant source of flux in the narrowband filter. This neglects a small continuum contribution to the total filter flux. However, we assume this continuum contribution is negligible due to the combination of the $(1+z)$ boosted equivalent widths (EWs) of the \Lya lines, the by-design narrow filter widths (NB816: $\sim$100~\AA{}, NB921: $\sim$130~\AA{}), and the strong \Lya breaks blue-ward of the \Lya lines. 

Following our approximation that the \Lya line accounts for all of the flux in the narrowband filters, for each LAE, we divide the corrected flux measurement by the narrowband filter throughput measured at the wavelength of the \Lya line peak. This accounts for the deviations of the filter transmission curves from ideal ``tophat'' curves, and it restores the more attenuated fluxes when the \Lya line lies near the edge of the narrowband filter. This correction also motivates our requirement that the LAEs in our final sample have \Lya lines that lie at wavelengths where there is $>50\%$ peak narrowband filter transmission. 

\begin{figure}[ht]
\includegraphics[width=\linewidth]{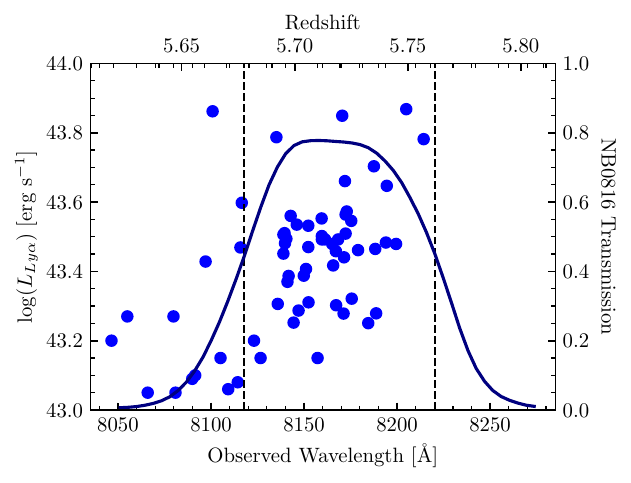}
\includegraphics[width=\linewidth]{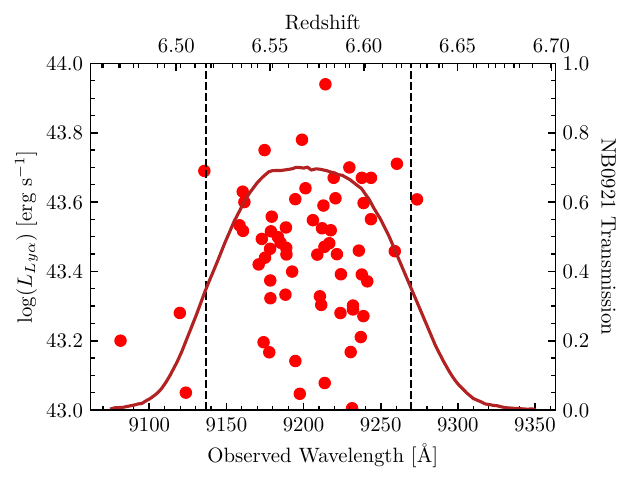}
\caption{Product of the filter transmission and the CCD quantum efficiency for the NB816/NB921 filter (blue/red curve). Blue/red circles show the redshift and observed-frame \Lya profile peak wavelength for each LAE at $z\sim5.7$/$z\sim6.6$. Vertical dashed lines show the redshift bounds adopted for this study, where the narrowband filter transmission is at $>$50\% of the peak transmission.}
\label{fig:LvLam}
\end{figure}

We use our spectroscopic redshifts to determine cosmological luminosity distances for each source, and hence \Lya luminosities. In Figure~\ref{fig:LvLam}, we plot the final \Lya luminosities as a function of redshift (and observed-frame wavelength) and overplot the NB816 and NB921 filter transmission curves. In Tables~\ref{tab:catalog57} and \ref{tab:catalog66}, we present our final catalogs of 49 $z=5.7$ LAEs and 56 $z=6.6$ LAEs.

\begin{figure}[ht]
\includegraphics[width=\linewidth]{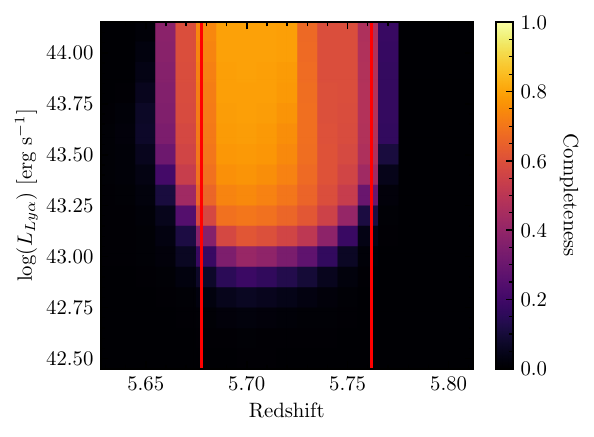}
\includegraphics[width=\linewidth]{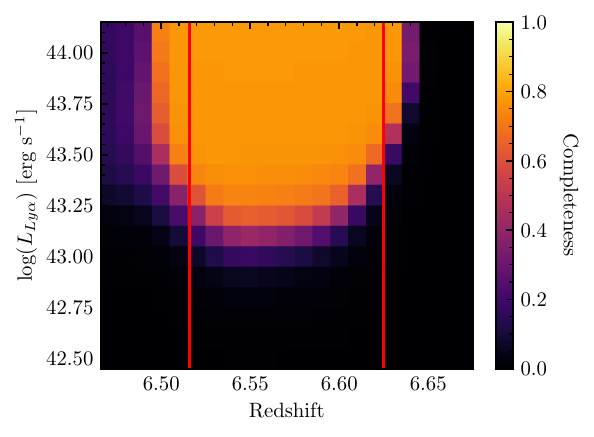}
\caption{Completeness (color scale) for $z=5.7$ LAEs (top) and $z=6.6$ LAEs (bottom) averaged over EW$_{\textrm{\Lya}}=25-75$~\AA{} and over the RA/Dec of the HEROES field plotted as a function of \Lya luminosity and redshift. We show our redshift bounds as vertical red lines in each panel. Both panels illustrate the dominant effect of the narrowband magnitude cut on completeness, as seen in the shapes of the narrowband filter transmission curves imprinted in the completeness data.}
\label{fig:LvZ}
\end{figure}

\begin{figure*}[ht]
\includegraphics[width=\linewidth]{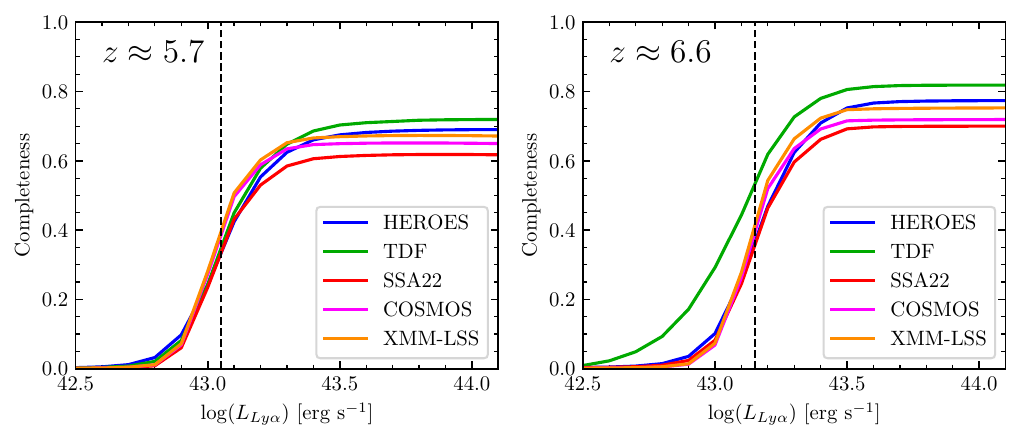}
\caption{Completeness versus \Lya luminosity for $z=5.7$ LAEs and $z=6.6$ LAEs in each of our fields, averaged over EW$_{\textrm{\Lya}}=25-75$~\AA{}, the RA/Dec of each field, and our redshift bounds. We plot vertical dashed lines at 
$L_{Ly\alpha}\sim10^{43.05}~\textrm{erg~s}^{-1}$ ($z=5.7$) and $L_{Ly\alpha}\sim10^{43.15}~\textrm{erg~s}^{-1}$ ($z=6.6$) to denote the theoretical cuts on $L_{Ly\alpha}$ imposed by our NB816$<24.5$ and NB921$<24.5$ selection cuts.}
\label{fig:CvL}
\end{figure*}

\section{Incompleteness Measurement}
\label{sec:completeness}
We adopt and revise the methodology of \cite{taylor21} to compute incompleteness corrections for our LAE samples. Briefly, to produce these corrections, we inject model LAEs into our existing HSC imaging data and determine the fraction of sources recovered and selected using the same methods and criteria that we used to select our LAE samples.

As in \citet{taylor21}, we use a simple LAE spectrum model consisting of a \Lya line modeled by a right triangle spanning 1215--1218~\AA{}, and a flat continuum (in $f_{\nu}$) redward of 1215~\AA. We then redshift this model spectrum and rescale the modeled flux to account for the redshift-determined luminosity distance. In this model, the \Lya luminosity, the continuum strength (set by the \Lya EW), and the redshift are the three tunable parameters that characterize a simulated spectrum. We simulate LAEs for all permutations of $\log_{10} \left( L_{\textrm{\Lya}}\right)$=42.5-44.0~ergs~s\per (in increments of 0.1~ergs~s\per), $z$=5.60--5.80, 6.45--6.70 (in increments of 0.01), and EW$_{\textrm{\Lya}}$=25--100~\AA{} (in increments of 25~\AA{}), resulting in a collection of 3375 unique model LAE spectra. For each of these, we multiply by the $g$, $r$, $i$, $z$, $y$, NB816, and NB921 filter transmission curves to determine the total flux that each filter would observe from the model spectrum. 

We next construct a simple 2D Gaussian model as a spatial model of an LAE in our imaging. We adopt a conservative FWHM of 0\farcs75, motivated by the primarily unresolved nature of the LAEs due to our seeing-limited ground-based HSC data. We determine the amplitude of the spatial LAE model, such that the model's integrated flux is equal to the synthetic observed fluxes derived from the spectral models. For each of the 3375 unique simulated spectra, we produce a corresponding set of flux-calibrated spatial model LAEs and inject them at random spatial positions in our HSC imaging at a density of $\sim$1~LAE per arcmin$^2$.

Next, we directly remeasure the injected synthetic LAEs using 2\farcs0 diameter aperture photometry. For each remeasured source, we apply our photometric color cuts (see Section~\ref{sec:observations}) and determine whether that source satisfies either our $z=5.7$ or $z=6.6$ LAE selection criteria. 
We repeat this process across all of our fields for each of the 3375 unique LAE model spectra. We record the fraction of synthetic LAEs based on each spectrum that we successfully remeasure and select as LAEs as the sample completeness for that model spectrum and its associated \Lya properties. 

Note that to save on computational time, we remeasure the injected source fluxes at the known injected positions instead of using a source detection algorithm. 
Due to our strong narrowband magnitude cut requirements (NB$<$24.5~magnitudes), all injected sources that satisfy these requirements are already detected at $>5\sigma$ in HEROES, where the median $5\sigma$ limiting magnitudes for NB921 and NB816 are both 24.4~magnitudes. Thus, running a detection algorithm---such as the \texttt{hscpipe} software \citep{bosch18} used in the original catalog constructions---with a much more sensitive multi-band detection threshold will recover all but the most blended synthetic sources. 

In Figure~\ref{fig:LvZ}, we show the results of these simulations for HEROES. At both redshifts, the narrowband magnitude cut is the dominant factor in the completeness. These selection cuts are apparent in the shapes of the narrowband transmission curves that appear in Figure~\ref{fig:LvZ}. However, in our final selection, our requirement that the LAEs lie at redshifts where the \Lya line lies within the FWHM of the filter transmission largely avoids areas with minimal to no completeness. 

For each field, we average the completeness over our redshift bounds. In Figure~\ref{fig:CvL}, we plot this overall completeness as a function of $L_{Ly\alpha}$. All fields at $z=5.7$ show a high degree of completeness ($>60\%$) at the bright end, and a sharp decline in completeness starting at $L_{Ly\alpha}\sim10^{43.05}~\textrm{erg s}^{-1}$, corresponding to the NB816$<24.5$ selection cut. Similarly, at $z=6.6$, all fields show an even higher degree of completeness ($>70\%$) at the bright end, with a sharp decline at $L_{Ly\alpha}\sim10^{43.15}~\textrm{erg s}^{-1}$, corresponding to the analogous NB921$<24.5$ selection cut. Of particular note is that the deeper NB921 and $z$ imaging in the TDF contribute to its superior 81\% completeness, and its deeper NB921$<25.0$ selection cut contributes to its slower decline to fainter luminosities. The bright end completeness at both redshifts is primarily limited by a combinations of spatial overlaps between real and simulated sources and detections of $g$-, $r$- or (for $z=6.6$) $i$-band emission during the completeness simulations. The $z=5.7$ completeness exhibits lower bright end completeness than at $z=6.6$, due to our $z-\textrm{NB816}>0.7$ selection cut. At low EW$\lesssim25$~\AA{}, the UV continuum is relatively strong compared to the \Lya line. As such, the UV continuum contributes significantly to the $z$-band flux, and can cause such an object to exhibit an insufficient $z-\textrm{NB816}$ narrowband excess, specifically when the \Lya line is offset from the peak of the NB816 filter transmission. This effect is clearly illustrated in Figure~\ref{fig:LvZ}, where the high-luminosity completeness at $z=5.7$ significantly changes as a function of redshift, while the $z=6.6$ high-luminosity completeness (where we do not apply analogous $y-$NB921 cut due to the limited depth of the $y$-band imaging) is quite constant between our redshift bounds. 

In all cases, it is clear that the shapes of the completeness curves are dominated by our selection cuts, indicating that our selection function and any resulting sample biases are well-characterized.

\begin{figure}[ht]
\includegraphics[width=\linewidth]{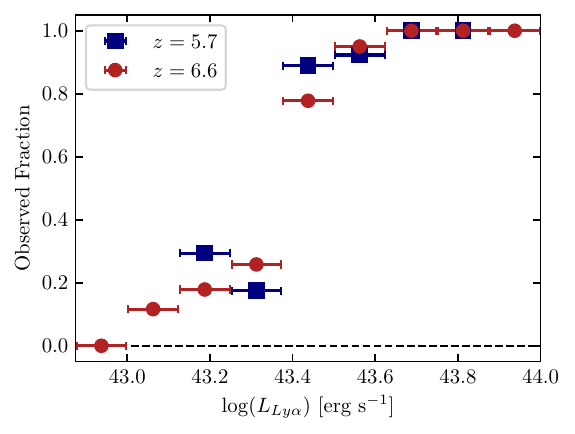}
\caption{Fraction of spectroscopically observed narrowband selected LAE candidates vs. \Lya luminosity for $z=5.7$ LAEs and $z=6.6$ LAEs across all of our targeted fields. 
}
\label{fig:obsfrac}
\end{figure}

\subsection{Spectroscopically Observed Fractions}
\label{sec:obsfrac}
As previously demonstrated in \citet{konno18,taylor20,taylor21}, spectroscopic confirmation of narrowband selected LAE candidates is essential to rejecting low-redshift contaminants (typically [O\,\textsc{iii}]$\lambda$5007 emitters) and artifacts from a photometric sample. In our Keck~II/DEIMOS observations (especially of the NEP and TDF fields), we have aimed to achieve and maintain a high fraction of spectroscopic completeness by observing sources from brightest to faintest in the NB816 and NB921 filters with exceptions made for the rare opportunities to include multiple LAE candidates on a single DEIMOS mask. In Figure~\ref{fig:obsfrac}, we plot the overall spectroscopically observed fraction as a function of estimated $L_{Ly\alpha}$.

At $z=5.7$, our spectroscopically observed fraction increases rapidly from $\approx$25\% at $L_{\textrm{\Lya}}\approx10^{43.25}~\textrm{erg~s}^{-1}$ to $\gtrsim90$\% at $L_{\textrm{\Lya}}\gtrsim10^{43.4}~\textrm{erg~s}^{-1}$.
This is consistent with our overall survey design, in which we first targeted the brightest candidates and have since targeted the more numerous fainter targets. The $z=6.6$ results extend to fainter luminosities (albeit with low observed fractions) due to the inclusion of fainter candidates in the TDF field that benefit from our targeted deeper NB921 and $z$ imaging.

\section{Luminosity Functions}
\label{sec:LFs}
Here we use our spectroscopic samples of LAEs at $z=5.7$ and $z=6.6$, together with our completeness simulations and spectroscopically observed fractions, to compute LAE LFs at each redshift. We compute the comoving volumes for each field assuming the flat cosmology given in Section~\ref{sec:intro}. We determine the redshift bounds for each comoving volume from where the narrowband filters are at $>$50\% of their peak transmission (see Figure~\ref{fig:LvLam}). We report the resulting comoving volumes in Table~\ref{tab:fields}. 

To construct the LFs for a single field, we use
\begin{equation}
\label{eq:lf}
\Phi(L)=\frac{N(L)}{C(L)\cdot V\cdot F(L) \cdot \Delta\log L} \,,
\end{equation}
where $N(L)$ is the number of spectroscopically identified LAEs within a given luminosity bin centered at \Lya luminosity $L$ with a bin width of $\Delta\log L$, $C(L)$ is the simulated completeness averaged over the redshift range of the LF and the luminosity interval, $V$ is the comoving volume, and $F(L)$ is the spectroscopically observed fraction in the same luminosity interval. 

Next, we generalize this expression to combine multiple fields, weighted by comoving volume, as
\begin{equation}
\label{eq:totallf}
\Phi(L)=\frac{\sum_i N_i(L)}{\sum_i \left[C_i(L)\cdot V_i\cdot F_i(L) \cdot \Delta\log L\right]} \,.
\end{equation}
Here, we sum the number of LAEs in a single field $N_i(L)$ within a given luminosity interval $\Delta\log L$ over multiple fields, and we divide by the sum of the products of the comoving volume, completeness, observed fraction, and luminosity interval to obtain the LF for our entire multi-field survey. To avoid fields and luminosity intervals with low spectroscopically observed fractions from significantly biasing the combined LF, we ignore contributions to the total LF from fields and luminosity intervals with an observed fraction $<$25\%. In these cases, the contribution from that field is removed from both the numerator and denominator of Equation~\ref{eq:totallf}.

We compute the uncertainty on the resulting combined LF, assuming that the variance in the observed number of LAEs dominates. We assume Poissonian uncertainties and use the formalism from \cite{gehrels86} to estimate these uncertainties. For the combined LFs, we use $\sum_i N_i(L)$ as the input $n$ to the \cite{gehrels86} equations to derive the uncertainty on the total number of observed LAEs in a luminosity interval, $\sigma_n(L)$. We then compute the uncertainty on the LF, $\sigma_{\Phi}(L)$, as
\begin{equation}
\label{eq:totallferr}
\sigma_{\Phi}(L)=\frac{\sigma_n(L)}{\sum_i \left[C_i(L)\cdot V_i\cdot F_i(L) \cdot \Delta\log L\right]} \,.
\end{equation}

\begin{deluxetable}{ccccc}[hbt]
\tablewidth{0pt}
\tablecaption{LAE Luminosity Function Data}
\label{tab:lf}
\tablehead{$\log (L)$ & $\Phi$ ($z=5.7$) & $\Phi$ ($z=6.6$) & N & N\cr
 [erg s$^{-1}$] & \multicolumn{2}{c}{[($\Delta\log L)^{-1}$~Mpc$^{-3}$]} & ($z=5.7$) & ($z=6.6$)}
\startdata
$43.250-43.375$ & $13.00^{+8.80}_{-5.62}$ & $12.48^{+6.72}_{-4.60}$ & 5 & 7 \\
$43.375-43.500$ & $5.18^{+1.53}_{-1.21}$ & $4.72^{+1.44}_{-1.13}$ & 18 & 17 \\
$43.500-43.625$ & $2.98^{+1.20}_{-0.89}$ & $2.77^{+0.96}_{-0.73}$ & 11 & 14 \\
$43.625-43.750$ & $0.84^{+0.82}_{-0.46}$ & $1.39^{+0.75}_{-0.51}$ & 3 & 7 \\
$43.750-43.875$ & $1.17^{+0.93}_{-0.56}$ & $0.38^{+0.50}_{-0.24}$ & 4 & 2 \\
$43.875-44.000$ & \nodata & $0.19^{+0.45}_{-0.16}$ & \nodata & 1 \\
\enddata
\end{deluxetable}

\begin{figure*}[ht]
\includegraphics[width=\linewidth]{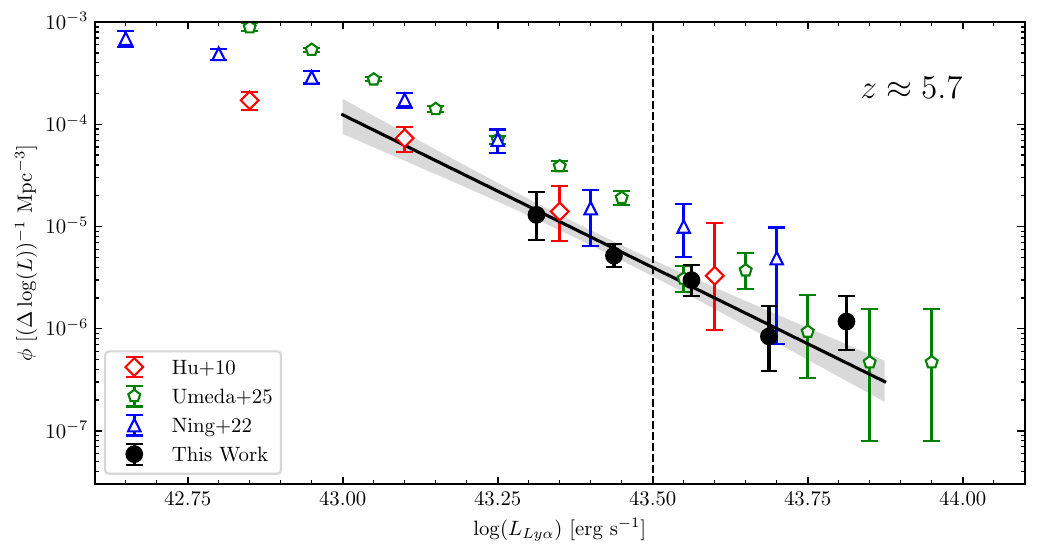}
\includegraphics[width=\linewidth]{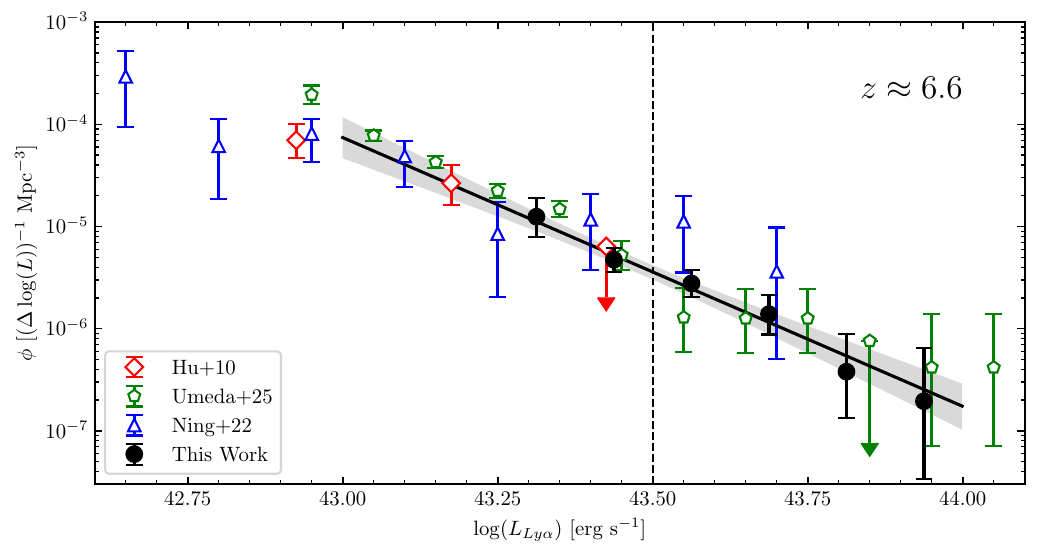}
\caption{
LF measurements for (top/bottom) the $z=5.7$/$z=6.6$ LAE samples from this work (black circles), \cite{hu10} (red diamonds), \cite{umeda25} (green pentagons), \cite{ning22} (blue triangles), and \cite{taylor21} (gray squares). 
We plot a vertical dashed line at $L_{\textrm{\Lya}}=10^{43.5}$~erg~s$^{-1}$ to separate the ultraluminous and sub-ultraluminous regions of the LFs. 
We plot the best fit power laws to the LFs from this work combined with the LFs from \cite{hu10} at $L_{\textrm{\Lya}}<10^{43.0}$~erg~s$^{-1}$ as solid black curves (with 1$\sigma$ uncertainties on the fits shown as the shaded gray regions).
}
\label{fig:66LF}
\end{figure*}

In Table~\ref{tab:lf}, we provide our resulting LAE LF data at $z=5.7$ and $z=6.6$. In Figure~\ref{fig:66LF}, we compare our LFs (black circles) with LFs from the literature. 
\cite{ning22} (blue triangles) used Magellan/M2FS to spectrsopically observe LAEs in the SXDS, A370, ECFFS, COSMOS, and SSA22 fields. Similarly, \cite{hu10} (red diamonds) used Keck/DEIMOS to observe LAES at fainter luminosities in the SSA22, A370, HDF-N, and SSA17 fields. \cite{ning22} and \cite{hu10} are purely spectroscopic studies, and are thus inherently contaminant free. \cite{umeda25} (green pentagons) used HSC-SSP narrowband photometric data to construct their LFs at $z=5.7$ and $z=6.6$. While they do not correct their LFs for low-redshift contaminants (estimated at $\sim14\%$ at $z=5.7$ and $\sim8\%$ at $z=6.6$ in their previous studies; i.e., \citealt{konno18,shibuya18}), they claim that their updated selection criteria successfully eliminate these previously identified interlopers. However, we note that the normalizations of the \cite{umeda25} LFs (especially at $z=5.7$) are significantly higher than the spectroscopic LFs from this work and \citep{hu10}, suggesting that there may still be some degree of contamination present in these purely photometric samples. 

At both $z=5.7$ and $z=6.6$, our LFs (especially when combined with the similarly selected and observed fainter LFs from \citealt{hu10}) strongly resemble featureless power laws, in contrast to the more commonly expected Schechter function \citep{schechter76} shapes typically observed in LFs. This is primarily due to the now well-established bright end excess of LAEs \citep{taylor21,ning22,umeda25}. Furthermore, no survey sufficiently samples the faint end of the LF ($L_{\textrm{\Lya}}\lesssim10^{42.75}$~erg~s$^{-1}$) to measure a faint end slope. Thus, we fit our LFs combined with the LFs from \cite{hu10} using a simple power law. 

At $z=5.7$, we use our LF at $L_{\textrm{\Lya}}>10^{43.375}$~erg~s$^{-1}$ and the LF from \cite{hu10} at $10^{43.0}<L_{\textrm{\Lya}}<10^{43.375}$~erg~s$^{-1}$. At $z=6.6$, we use our full LF and the LF from \cite{hu10} at $10^{43.0}<L_{\textrm{\Lya}}<10^{43.25}$~erg~s$^{-1}$ to best sample the bright end slope. We fit the bright end slope using a simple power law model:
\begin{equation}
\label{eq:powerlaw}
\Phi(L)=\Phi_{L=10^{43.5}}\cdot \left(\frac{L}{10^{43.5}}\right)^{\beta+1} \,.
\end{equation} 
Here, $\Phi_{L=10^{43.5}}$ is the normalization of the LF at $L=10^{43.5}~\textrm{erg s}^{-1}$, and $\beta$ is the bright end slope. We use a Markov-Chain-Monte-Carlo (MCMC) fitter \citep[\texttt{emcee};][]{emcee} to fit this power law to our data using a flat linear prior for $\beta$, a flat $\log$ prior for $\Phi_{L=10^{43.5}}$, and a least-squares likelihood function. 

In Figure~\ref{fig:66LF}, we plot our best fit power laws and $1\sigma$ uncertainties as black lines and gray shaded regions. We derive bright end slopes of $-3.98^{+0.41}_{-0.36}$ at $z=5.7$ and $-3.62^{+0.39}_{-0.40}$ at $z=6.6$. These spectroscopic sample bright end slopes compare favorably (nearly within uncertainties) to the bright end slopes fit by \cite{umeda25} using double power law models (bright and faint end slopes) to their photometric LFs with bright end slopes of $-4.20^{+0.12}_{-0.13}$ at $z=5.7$ and $-4.17^{+0.30}_{-0.59}$ at $z=6.6$.

Due to our large 67.8~deg$^2$ combined survey area, our LFs (and those of \citealt{umeda25}, using the combined HSC-SSP deep/ultra-deep fields) are able to probe the brightest yet-observed end of the \Lya LFs. Using our power-law fit to our $z=6.6$ \Lya LF, we derive an estimator of sky area per LAE at even brighter luminosities: 
\begin{multline}
\log (A_{z=6.6})= \\ 2.617\log(L_{\textrm{\Lya}})-\log(\Delta\log(L_{\textrm{\Lya}}))-114.3587.
\label{eq:area66}
\end{multline}
Here, $A_{z=6.6}$ is the area in square degrees necessary to find a single $z=6.6$ LAE with in a bin of width $\Delta\log(L_{\textrm{\Lya}})$ centered at a \Lya luminosity of $L_{\textrm{\Lya}}$. We recompute the same estimate for $z=5.7$:
\begin{multline}
\log (A_{z=5.7})= \\ 3.003\log(L_{\textrm{\Lya}})-\log(\Delta\log(L_{\textrm{\Lya}}))-131.1978.
\label{eq:area57}
\end{multline}
Using Equation~\ref{eq:area66}, we estimate that finding a single $z=6.6$ at $L_{\textrm{\Lya}}=10^{44.0}-10^{44.125}$~erg~s$^{-1}$ would require 74~deg$^{2}$ of survey area. To advance an additional bin brighter to $L_{\textrm{\Lya}}=10^{44.125}-10^{44.25}$~erg~s$^{-1}$ would require a survey area of 160~deg$^{2}$. Alternatively, combining these bins results in an estimated area of 54~deg$^{2}$. This area is already covered both by this program and by \cite{umeda25}, who report a single photometric object at $L_{\textrm{\Lya}}\approx10^{44.05}$. However, to truly expand the $z=6.6$ LF to even brighter luminosities, exponentially larger narrowband surveys are clearly necessary. Such surveys would not require significant narrowband depth (a $z=6.6$ LAE with $L_{\textrm{\Lya}}=10^{44.25}$~erg~s$^{-1}$ would produce NB921=21.7~magnitudes). Our proven strategy of combining Subaru/HSC imaging and Keck/DEIMOS (or Keck/KCWI) spectroscopic followup could efficiently accomplish this, especially if the new narrowband imaging is obtained in fields already covered in \textit{g, r, i, z,} and \textit{y} broadbands by the HSC-SSP-Wide program \citep{aihara18,aihara18b,aihara19,aihara22}.

\begin{figure*}[ht]
\includegraphics[width=\linewidth]{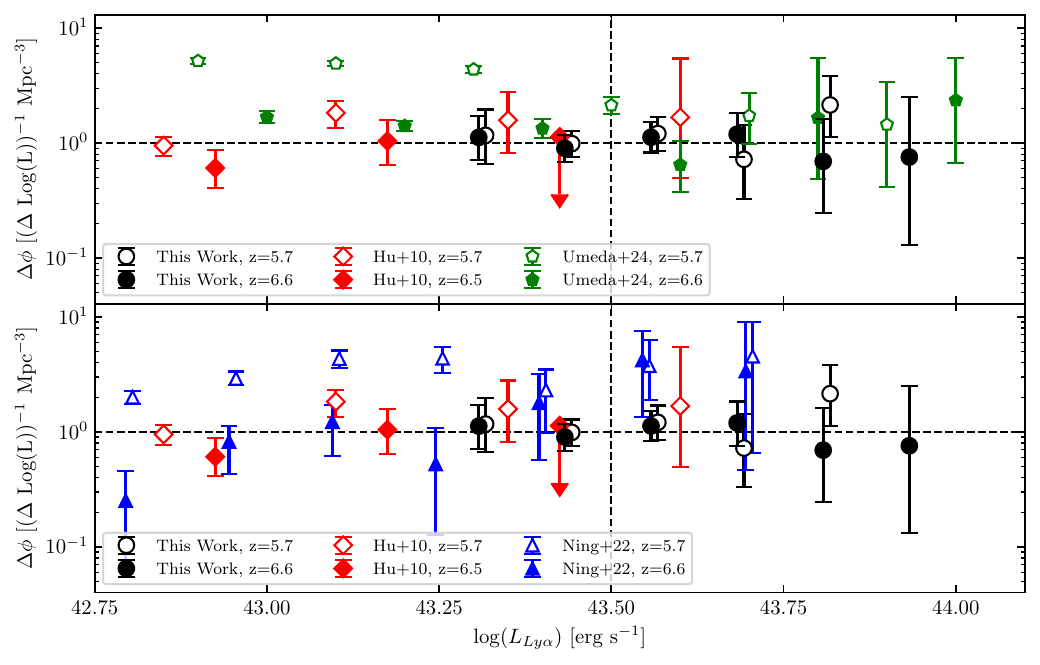}
\caption{Evolution of the LAE LF over the redshift range $z=6.6$ to $z=5.7$ for the \citet{ning22} and \citet{umeda25} surveys, shown relative to ours and the \cite{hu10} LFs. In each panel, we divide by the power law fit to ours plus the \citet{hu10} $z=6.6$ LF to better examine the evolution in both the ultraluminous and sub-ultraluminous regimes. We again plot a vertical dashed line at $L_{\textrm{\Lya}}=10^{43.5}$~erg~s$^{-1}$ to separate these regimes. Note that we plot the \cite{umeda25} and \cite{ning22} points on separate subplots and introduce small $\log L_{\textrm{\Lya}}=\pm0.005$ offsets between the $z=5.7$ and $z=6.6$ points for visual clarity. We also re-bin the \cite{umeda25} data by a factor of two to reduce the bright-end error bars and improve visual clarity.
}
\label{fig:5766}
\end{figure*}

In Figure~\ref{fig:5766}, we divide both our LF and the literature LFs by the power law fit to our $z=6.6$ LF to best illustrate the evolution (or lack thereof) of the LFs over the redshift range $z=6.6$ to $z=5.7$. In our comparisons, we use the \cite{hu10} LFs as faint end extensions of our own work, as they were measured and constructed in a similar manner. 

Below the ultraluminous threshold, there is a separation of $\sim$0.6~dex between the \cite{umeda25} $z=6.6$ and $z=5.7$ photometric LFs, but they converge to within their error bars by $L_{\textrm{\Lya}}\approx10^{43.75}~\textrm{erg~s}^{-1}$. Meanwhile, the \cite{ning22} spectroscopic LFs show the same $\sim$0.6~dex separation below $L_{\textrm{\Lya}}\approx10^{43.25}~\textrm{erg~s}^{-1}$, but then they converge much more quickly at $L_{\textrm{\Lya}}\approx10^{43.4}~\textrm{erg~s}^{-1}$. 
Our LFs also converge by $L_{\textrm{\Lya}}\approx10^{43.3}~\textrm{erg~s}^{-1}$. Notably, while our LFs strongly confirm the convergence of the $z=5.7$ and $z=6.6$ LFs at the ultraluminous end, the \cite{hu10} LFs at the faint end show a much less significant evolution than either the \cite{ning22} or \cite{umeda25} LFs. This may be due to the relatively small probed area in the \cite{hu10} LFs ($\sim1.2$~deg$^{-1}$), but as \cite{hu10} observed multiple fields to construct their LFs, the effects of cosmic variance should be reduced over what might be expected from an equivalent continuous area. Regardless, while our LFs do not yet probe faint enough to demonstrate evolution in the sub-ultraluminous regime, they definitively demonstrate the bright end convergence. Moreover, as our LFs show convergence as faint as $L_{\textrm{\Lya}}\approx10^{43.3}~\textrm{erg~s}$, we suggest (as also motivated in \citealt{songaila22} from line width measurements) the the ultraluminous threshold be reduced from $L_{\textrm{\Lya}}=10^{43.5}~\textrm{erg~s}$ to $L_{\textrm{\Lya}}=10^{43.25}~\textrm{erg~s}$ in future studies.

This convergence at the bright end is widely observed and may indicate the presence of ionized bubbles around the most luminous LAEs at $z=6.6$ \citep{matthee15,santos16,taylor20,taylor21,ning22}. Such bubbles would permit increased \Lya transmission through the otherwise neutral IGM, as the \Lya line has a greater distance over which to redshift off of resonance before encountering the neutral IGM. As the IGM is primarily ionized at $z=5.7$ \citep[e.g.,][]{finkelstein19,Smith2022}, the $z=5.7$ LF shows a strong evolution---especially at the faint end---relative to the $z=6.6$ LF due to the elimination of the IGM attenuation. However, because the attenuation of the bright end of the $z=6.6$ LF is reduced by the effects of ionized bubbles, this evolution is minimized in the ultraluminous regime, matching our observations. Notably, \cite{songaila22,songaila24}, using the same dataset that we use in this work, examined the evolution of the \Lya linewidths as a function of redshift and \Lya luminosity and observed the same effect. 

The bright end convergence exhibited in our LFs significantly supports the ionized bubble model, suggesting that even if reionization were primarily driven by faint galaxies \citep{finkelsteinbagley22}, the most luminous sources also played a significant role in reionization. 


\section{Summary}\label{sec:summary}

We summarize the work as follows:

\begin{enumerate}
\item{We examined two samples of LAEs from \cite{songaila24}, one at $z=5.7$ and one at $z=6.6$, which were selected using Subaru/HSC broadband and narrowband photometry and followed up spectroscopically with Keck II/DEIMOS.}

\item{We applied photometric and spectroscopic cuts to these samples to produce uniformly selected samples of 49 $z=5.7$ and 56 $z=6.6$ LAEs.}

\item{We computed a robust completeness correction using a source-injection simulation and quantified the spectroscopic completeness of our samples.}

\item{We computed ULLAE LFs at $z=5.7$ and $z=6.6$ and found excellent agreement with current spectroscopic \citep{ning22} and photometric \citep{umeda25} LFs from the literature.}

\item{We tested our LFs for evolution over the redshift range $z=6.6$ to $z=5.7$. We found a strong convergence of the LFs at $L_{\textrm{\Lya}}\gtrsim 10^{43.4}~\textrm{erg~s}^{-1}$, which may provide evidence for the presence of ionized bubbles around the ULLAEs. This suggests that ULLAEs may have played a significant role in cosmic reionization.}
\end{enumerate}

\section*{Acknowledgments}
We gratefully acknowledge support for this research from the UT-Austin College of Natural Sciences (A.J.T.) a Wisconsin Space Grant Consortium Graduate and Professional Research Fellowship (A.J.T.), a Sigma Xi Grant in Aid of Research (A.J.T.), NSF grants AST-1715145 (A.J.B) and AST-1716093 (E.M.H., A.S.), and a Kellett Mid-Career Award and a WARF Named Professorship from the University of Wisconsin--Madison Office of the Vice Chancellor for Research and Graduate Education with funding from the Wisconsin Alumni Research Foundation (A.J.B.). 

This paper is based in part on data from the Subaru Telescope. The Hyper Suprime-Cam (HSC) collaboration includes the astronomical communities of Japan and Taiwan, and Princeton University. The HSC instrumentation and software were developed by the National Astronomical Observatory of Japan (NAOJ), the Kavli Institute for the Physics and Mathematics of the Universe (Kavli IPMU), the University of Tokyo, the High Energy Accelerator Research Organization (KEK), the Academia Sinica Institute for Astronomy and Astrophysics in Taiwan (ASIAA), and Princeton University. Funding was contributed by the FIRST program from Japanese Cabinet Office, the Ministry of Education, Culture, Sports, Science and Technology (MEXT), the Japan Society for the Promotion of Science (JSPS), Japan Science and Technology Agency (JST), the Toray Science Foundation, NAOJ, Kavli IPMU, KEK, ASIAA, and Princeton University. 

This paper also makes use of data collected at the Subaru Telescope and retrieved from the HSC data archive system, which is operated by the Subaru Telescope and Astronomy Data Center at National Astronomical Observatory of Japan. Data analysis was in part carried out with the cooperation of Center for Computational Astrophysics, National Astronomical Observatory of Japan.

This paper is based in part on data collected from the Keck~II Telescope. The W.~M.~Keck Observatory is operated as a scientific partnership among the California Institute of Technology, the University of California, and NASA, and was made possible by the generous financial support of the W.~M.~Keck Foundation.

The authors wish to recognize and acknowledge the very significant cultural role and reverence that the summit of Maunakea has always had within the indigenous Hawaiian community. We are most fortunate to have the opportunity to conduct observations from this mountain.

\facilities{JWST}

\software{astropy: \cite{astropy:2013,astropy:2018,astropy22}, emcee: \cite{emcee}}

\bibliographystyle{aasjournalv7.bst}
\bibliography{ref1.bib}

\appendix
\section{Tables of Sources}
\startlongtable
\begin{deluxetable*}{ccccc}
\renewcommand\baselinestretch{1.0}
\tablewidth{0pt}
\tablecaption{Properties of the $z=5.7$ Spectroscopically Observed Sample}
\label{tab:catalog57}
\tablehead{Source & R.A. & Decl. & Redshift & $\log (L_{\textrm{\Lya}})$ \\ 
 & (deg) & (deg) & & (erg~s$^{-1}$)}
\startdata
XMM-LSS\_33.869-4.815 & 33.869137 & -4.815254 & 5.7166 & 43.48 \\
XMM-LSS\_34.103-4.921 & 34.102951 & -4.921283 & 5.7050 & 43.41 \\
XMM-LSS\_34.350-5.552 & 34.350140 & -5.552704 & 5.7060 & 43.53 \\
XMM-LSS\_34.651-5.591 & 34.651592 & -5.591156 & 5.6973 & 43.39 \\
XMM-LSS\_35.355-3.676 & 35.355396 & -3.676839 & 5.6982 & 43.56 \\
XMM-LSS\_35.496-4.975 & 35.496708 & -4.975223 & 5.6968 & 43.37 \\
XMM-LSS\_35.941-3.904 & 35.941269 & -3.904916 & 5.7360 & 43.28 \\
NEP\_259.37+66.385 & 259.379425 & 66.385239 & 5.7009 & 43.53 \\
NEP\_261.20+65.015 & 261.203156 & 65.015106 & 5.7403 & 43.48 \\
NEP\_262.36+68.034 & 262.364502 & 68.034317 & 5.7407 & 43.65 \\
NEP\_263.36+68.198 & 263.364624 & 68.198364 & 5.6957 & 43.48 \\
NEP\_263.44+66.333 & 263.444122 & 66.332985 & 5.7183 & 43.30 \\
NEP\_265.63+68.497 & 265.636658 & 68.497589 & 5.7182 & 43.46 \\
NEP\_266.47+67.920 & 266.475281 & 67.920891 & 5.7040 & 43.39 \\
NEP\_267.68+66.890 & 267.680878 & 66.890022 & 5.7230 & 43.57 \\
NEP\_267.78+66.727 & 267.787964 & 66.727745 & 5.6963 & 43.49 \\
NEP\_267.89+66.629 & 267.892090 & 66.629250 & 5.6919 & 43.79 \\
NEP\_268.73+64.312 & 268.735291 & 64.312477 & 5.7280 & 43.46 \\
NEP\_268.74+64.242 & 268.741760 & 64.242134 & 5.6950 & 43.45 \\
NEP\_270.16+68.816 & 270.159821 & 68.815956 & 5.6950 & 43.51 \\
NEP\_270.65+67.816 & 270.656860 & 67.815994 & 5.7192 & 43.49 \\
NEP\_270.96+68.079 & 270.963257 & 68.079491 & 5.7250 & 43.55 \\
NEP\_271.01+67.018 & 271.016113 & 67.017967 & 5.7134 & 43.49 \\
NEP\_271.34+67.920 & 271.339874 & 67.920433 & 5.7210 & 43.85 \\
NEP\_271.72+64.856 & 271.723083 & 64.856842 & 5.7061 & 43.31 \\
NEP\_271.95+64.922 & 271.953430 & 64.922447 & 5.7225 & 43.56 \\
NEP\_272.35+64.833 & 272.357910 & 64.833618 & 5.7017 & 43.29 \\
NEP\_272.41+66.774 & 272.416077 & 66.774391 & 5.7448 & 43.48 \\
NEP\_272.58+66.690 & 272.584717 & 66.690491 & 5.7350 & 43.70 \\
NEP\_273.02+68.850 & 273.023163 & 68.850861 & 5.7222 & 43.66 \\
NEP\_273.24+69.008 & 273.247772 & 69.008827 & 5.6995 & 43.25 \\
NEP\_274.18+66.973 & 274.185120 & 66.973175 & 5.7570 & 43.78 \\
NEP\_274.55+67.849 & 274.552979 & 67.849693 & 5.7225 & 43.51 \\
NEP\_274.67+64.777 & 274.674500 & 64.777328 & 5.7218 & 43.44 \\
NEP\_274.79+66.034 & 274.794495 & 66.034172 & 5.7252 & 43.32 \\
NEP\_275.87+64.566 & 275.870697 & 64.565979 & 5.7356 & 43.46 \\
NEP\_276.02+64.449 & 276.026886 & 64.449387 & 5.7170 & 43.42 \\
NEP\_276.66+67.531 & 276.664093 & 67.531830 & 5.7120 & 43.50 \\
SSA22\_333.67-0.269 & 333.672974 & -0.269409 & 5.7119 & 43.55 \\
SSA22\_333.80-0.494 & 333.807037 & -0.494586 & 5.6955 & 43.51 \\
SSA22\_334.22+0.0941 & 334.228882 & 0.094081 & 5.6849 & 43.15 \\
SSA22\_334.47+1.1150 & 334.471222 & 1.115050 & 5.7216 & 43.28 \\
SSA22\_334.53+1.1232 & 334.537659 & 1.123230 & 5.7493 & 43.87 \\
SSA22\_334.54+0.0835 & 334.547852 & 0.083564 & 5.7101 & 43.15 \\
SSA22\_334.63+0.4790 & 334.636169 & 0.479069 & 5.6820 & 43.20 \\
SSA22\_334.79-0.510 & 334.799103 & -0.510763 & 5.7060 & 43.47 \\
SSA22\_334.87-0.936 & 334.873474 & -0.936297 & 5.6925 & 43.31 \\
SSA22\_334.89-0.781 & 334.896484 & -0.781477 & 5.7120 & 43.49 \\
SSA22\_335.85-0.693 & 335.854645 & -0.693872 & 5.7325 & 43.25 \\
\enddata
\end{deluxetable*}

\startlongtable
\begin{deluxetable*}{ccccc}
\renewcommand\baselinestretch{1.0}
\tablewidth{0pt}
\tablecaption{Properties of the $z=6.6$ Spectroscopically Observed Sample}
\label{tab:catalog66}
\tablehead{Source & R.A. & Decl. & Redshift & $\log (L_{\textrm{\Lya}})$ \\ 
 & (deg) & (deg) & & (erg~s$^{-1}$)}
\startdata
XMM-LSS\_35.580-3.535 & 35.580456 & -3.535472 & 6.5556 & 43.48 \\
XMM-LSS\_35.903-3.933 & 35.903450 & -3.933697 & 6.5473 & 43.44 \\
XMM-LSS\_36.098-4.009 & 36.098125 & -4.009507 & 6.5634 & 43.61 \\
COSMOS\_150.24+1.8041 & 150.241776 & 1.804108 & 6.6038 & 43.67 \\
COSMOS\_150.35+2.5292 & 150.353363 & 2.529239 & 6.5439 & 43.42 \\
COSMOS\_150.64+2.2038 & 150.647415 & 2.203886 & 6.5922 & 43.70 \\
NEP\_259.78+65.388 & 259.788574 & 65.388054 & 6.5773 & 43.30 \\
NEP\_259.87+66.158 & 259.875488 & 66.158943 & 6.5878 & 43.39 \\
NEP\_259.91+65.381 & 259.913147 & 65.381004 & 6.5752 & 43.45 \\
NEP\_260.45+66.052 & 260.459473 & 66.052803 & 6.5984 & 43.21 \\
NEP\_260.79+66.069 & 260.791626 & 66.069168 & 6.5942 & 43.29 \\
NEP\_260.80+65.373 & 260.802582 & 65.373360 & 6.5815 & 43.48 \\
NEP\_260.88+65.407 & 260.880615 & 65.407745 & 6.5465 & 43.20 \\
NEP\_261.04+65.029 & 261.042938 & 65.029694 & 6.5766 & 43.33 \\
NEP\_261.41+65.315 & 261.418152 & 65.315002 & 6.5875 & 43.28 \\
NEP\_261.46+65.413 & 261.467468 & 65.413277 & 6.5930 & 43.17 \\
NEP\_261.50+65.975 & 261.501465 & 65.975975 & 6.5942 & 43.30 \\
NEP\_261.51+65.963 & 261.511658 & 65.963448 & 6.5635 & 43.14 \\
NEP\_261.67+65.837 & 261.670258 & 65.837273 & 6.5496 & 43.17 \\
NEP\_261.70+65.796 & 261.708588 & 65.796097 & 6.5617 & 43.40 \\
NEP\_262.02+66.044 & 262.029022 & 66.044090 & 6.5997 & 43.27 \\
NEP\_262.30+65.599 & 262.308411 & 65.599686 & 6.5689 & 43.64 \\
NEP\_262.44+65.180 & 262.442963 & 65.180443 & 6.5670 & 43.78 \\
NEP\_263.04+65.543 & 263.043030 & 65.543968 & 6.6018 & 43.37 \\
NEP\_263.44+66.975 & 263.444092 & 66.975113 & 6.6176 & 43.71 \\
NEP\_263.49+67.467 & 263.493011 & 67.467751 & 6.5584 & 43.53 \\
NEP\_263.56+67.551 & 263.561401 & 67.551552 & 6.5502 & 43.32 \\
NEP\_263.58+67.668 & 263.583984 & 67.668724 & 6.5586 & 43.47 \\
NEP\_263.61+67.593 & 263.614899 & 67.593971 & 6.5839 & 43.67 \\
NEP\_265.22+65.510 & 265.224365 & 65.510391 & 6.5989 & 43.67 \\
NEP\_266.12+68.974 & 266.129181 & 68.974747 & 6.5849 & 43.61 \\
NEP\_266.14+64.722 & 266.143372 & 64.722610 & 6.5790 & 43.47 \\
NEP\_267.39+65.911 & 267.391052 & 65.911667 & 6.5355 & 43.52 \\
NEP\_267.42+68.907 & 267.427002 & 68.907410 & 6.5500 & 43.47 \\
NEP\_268.06+65.406 & 268.061768 & 65.406746 & 6.5856 & 43.45 \\
NEP\_268.29+65.109 & 268.292114 & 65.109581 & 6.5471 & 43.75 \\
NEP\_268.36+68.317 & 268.362701 & 68.317719 & 6.5973 & 43.46 \\
NEP\_269.68+65.944 & 269.689636 & 65.944748 & 6.5363 & 43.60 \\
NEP\_270.76+67.261 & 270.768890 & 67.261864 & 6.5456 & 43.49 \\
NEP\_271.02+64.532 & 271.020752 & 64.532166 & 6.5989 & 43.39 \\
NEP\_271.19+64.343 & 271.192322 & 64.343971 & 6.5587 & 43.45 \\
NEP\_271.41+68.226 & 271.409760 & 68.226112 & 6.6165 & 43.46 \\
NEP\_271.92+64.798 & 271.923706 & 64.798882 & 6.5501 & 43.37 \\
NEP\_272.30+67.434 & 272.308044 & 67.434807 & 6.5823 & 43.52 \\
NEP\_272.55+67.261 & 272.558807 & 67.261765 & 6.5998 & 43.60 \\
NEP\_272.66+67.386 & 272.661041 & 67.386055 & 6.5784 & 43.59 \\
NEP\_273.37+65.414 & 273.379303 & 65.414803 & 6.5509 & 43.56 \\
NEP\_273.41+68.092 & 273.411896 & 68.092964 & 6.5777 & 43.52 \\
NEP\_273.42+67.516 & 273.420776 & 67.516861 & 6.5542 & 43.50 \\
NEP\_273.73+65.285 & 273.738373 & 65.285995 & 6.5795 & 43.94 \\
NEP\_274.70+64.570 & 274.702637 & 64.570213 & 6.5728 & 43.55 \\
NEP\_274.90+66.126 & 274.900055 & 66.126717 & 6.5337 & 43.53 \\
NEP\_275.76+68.555 & 275.763885 & 68.555733 & 6.5504 & 43.52 \\
NEP\_276.23+67.606 & 276.234283 & 67.606705 & 6.5354 & 43.63 \\
SSA22\_333.68-0.572 & 333.681122 & -0.572996 & 6.5582 & 43.33 \\
SSA22\_334.90-0.071 & 334.908295 & -0.071056 & 6.6038 & 43.55 \\
\enddata
\end{deluxetable*}

\end{document}